\documentclass[epj]{webofc}
\usepackage[varg]{txfonts}   
\usepackage{subfigure} 
\usepackage{amssymb,amsmath,graphicx,xcolor,cancel,hyperref,slashed}

\wocname{EPJ Web of Conferences}
\woctitle{CONF12}
\newcommand\Dslash{\slashed D}
\newcommand{\tsep}{\mathop{t_{\rm sep}}\nolimits}
\newcommand{\tsepi}{\mathop{t_{\rm sep} \to \infty}\nolimits}

\newcommand{\Tr}{\mathop{\rm Tr}\nolimits}

\newcommand{\CPV}{$\cancel{\text{CP}}$}

\begin{document}
\selectlanguage{english}
\title{The Contribution of Novel CP Violating Operators to the nEDM using Lattice QCD}
%
%

\author{Rajan Gupta\inst{1}\fnsep\thanks{Speaker. \email{rajan@lanl.gov}. Los Alamos Report LA-UR-16-29579} \and
        Tanmoy Bhattacharya\inst{1} \and
        Vicenzo Cirigliano\inst{1}  \and
        Boram Yoon\inst{1}  
}

\institute{Los Alamos National Laboratory, Theoretical Division, T-2, Los Alamos, NM 87545}

\abstract{In this talk, we motivate the calculation of the matrix
  elements of novel CP violating operators, the quark EDM and the
  quark chromo EDM operators, within the nucleon state using lattice
  QCD. These matrix elements, combined with the bound on the neutron
  EDM, would provide stringent constraints on beyond the standard
  model physics, especially as the next generation of neutron EDM
  experiments reduce the current bound.  We then present our lattice
  strategy for the calculation of these matrix elements, in particular
  we describe the use of the Schrodinger source method to reduce the
  calculation of the 4-point to 3-point functions needed to evaluate
  the quark chromo EDM contribution. We end with a status report on
  the quality of the signal obtained in the lattice calculations of
  the connected contributions to the quark chromo EDM operator and the 
  pseudoscalar operator it mixes with under renormalization.  }
\maketitle
\section{Introduction: Baryogenesis and the need for novel CP violation}
\label{intro}

The observed universe has \(6.1^{+0.3}_{-0.2}\times 10^{-10}\)
baryons for every black body photon~\cite{Bennett:2003bz}, whereas in a baryon
symmetric universe, we expect no more that about \(10^{-20}\) baryons
for every photon~\cite{Kolb:1990vq}.  It is difficult to include such a
large excess of baryons as an initial condition in an inflationary
cosmological scenario~\cite{Coppi:2004za}.  The way out of the impasse
lies in generating the baryon excess dynamically during the evolution
of the universe.  

In the early history of the universe, if the matter-antimatter
symmetry was broken post inflation and reheating, then one is faced
with Sakharov's three necessary conditions~\cite{Sakharov:1967dj}: the
process has to violate baryon number, evolution has to occur out of
equilibrium, and charge-conjugation and T (or equivalently CP if CPT
remains a good symmetry) invariance has to be violated.

CP violation (\CPV) exists in the standard model (SM) of particle
interactions due to a phase in the Cabibo-Kobayashi-Maskawa (CKM)
quark mixing matrix~\cite{Kobayashi:1973fv}, and possibly by a similar
phase in the leptonic sector, given that the neutrinos are not
massless~\cite{Nunokawa:2007qh}.  \CPV\  arising from the CKM
matrix in the SM contributes at $O(10^{-32})$ e-cm, much smaller than
the current experimental bound $d_n < 2.9 \times 10^{-26}$ e-cm. Thus
the nEDM puts no constrain on the SM and the strength of the \CPV\ in
the CKM matrix is much too small to explain Baryogenesis.

In principle, the SM has an additional source of \CPV\  arising
from the effect of QCD instantons.  The presence of these finite
action non-perturbative configurations in a non-Abelian theory often
leads to inequivalent quantum theories defined over various
`$\Theta$'-vacua~\cite{Dolgov:1991fr}. Because of asymptotic freedom, all
non-perturbative configurations including instantons are strongly
suppressed at high temperatures where baryon number violating
processes occur. Because of this, \CPV\  due to such vacuum
effects do not lead to appreciable baryon number production. In short, 
additional much larger \CPV\  is needed from physics beyond the SM (BSM).

To determine whether such additional \CPV\  exists, a very promising approach
is to measure the static electric dipole moments of elementary
particles, atoms and molecules, which are necessarily proportional to their spin. Since
under time-reversal the direction of spin reverses but the electric
dipole moment does not, a non-zero measurement confirms T violation or equivalently \CPV.
Of the elementary particles, atoms and nuclei that are being
investigated, the electric dipole moment of the neutron (nEDM) is the
laboratory where lattice QCD can provide the theoretical part of the
calculation needed to ``connect'' the experimental bound (value) on the
nEDM to the strength of \CPV\  in a given BSM theory. 

Most extensions of the SM have new sources of \CPV.  Each of these
contributes to the nEDM and for some models it can be as large as $
10^{-26}$ e-cm.  Planned experiments are aiming to reduce the bound on
$d_n < 2.9 \times 10^{-26}$ e-cm to $d_n \lesssim 3 \times 10^{-28}$ e-cm.
The strategy for finding out which class of BSM theories are viable
candidates is as follows: As the bound on the nEDM is lowered in
current and planned experiments, BSM theories with \CPV\ giving rise
to a nEDM larger than this bound get ruled out provided the
``connection'' between the bound and the couplings is known
sufficiently accurately.  This ``connection'' is the matrix elements
of the novel CP violating interactions within the neutron state that
simulations of lattice QCD are gearing up to provide.

\section{QCD Lagrangian and the EM current in the presence of CP Violation}
\label{sec:Lagrangian}

Many candidate BSM theories have been proposed by theorists.  While
the true BSM theory is not known, one can write down all possible \CPV\ 
interactions at the energy scale of hadronic matter ($\sim$ few~GeV) in terms
of quark and gluon fields based on symmetry and organized by their
dimension; operators with higher dimension being, in general,
suppressed. At the lowest dimension five, there are two leading
operators called the quark EDM (qEDM) and the quark chromo EDM (CEDM).  The QCD
Lagrangian in the presence of the $\Theta$-term, the qEDM and the
CEDM operators is 

{\begin{equation}
{\cal L}_{\rm QCD}  \mathbin{{\longrightarrow}} {\cal L}_{\rm QCD}^{\cancel{\text{CP}}} =  {\cal L}_{\rm QCD}  + 
   i \Theta G_{\mu\nu} {\tilde G_{\mu\nu}} \ + 
   \left. i \sum_q d_q^\gamma \overline{q} \Sigma^{\mu\nu} {\tilde F_{\mu\nu}} q \right. + 
   \left. i \sum_q d_q^G \overline{q} \Sigma^{\mu\nu} {\tilde G_{\mu\nu}} q \right.\!\!\!\!
\label{eq:Lcpv}
\end{equation}}%
where the sum is over all the quark flavors, $\Sigma^{\mu\nu} =
[\gamma^\mu,\gamma^\nu]/2$, and $F_{\mu\nu}$ and $G_{\mu\nu}$ are the
electromagnetic and chromo field strength tensors.  The couplings
$d_{q}^{\gamma}$ are the qEDMs and the $d_{q}^{G}$ are the CEDMs.
They parameterize new \CPV\ in BSM theories. The goal is to constrain
BSM models by bounding these couplings using the experimental bound on
the neutron EDM and lattice QCD calculations of the matrix elements of
the corresponding operators. In other words, the matrix elements of
the electromagnetic current $J^{\rm EM}_\mu$ between neutron states in
the presence of CP violation provides the ``connection'' between the
BSM couplings and the nEDM.

\begin{center}
\begin{figure}[t]
\renewcommand{\tabcolsep}{0.0375\textwidth}
\begin{tabular}{cc}
\fbox{QEDM operator}%
&%
\fbox{$\Theta$-term}%
\\
\includegraphics[width=0.2\textwidth]{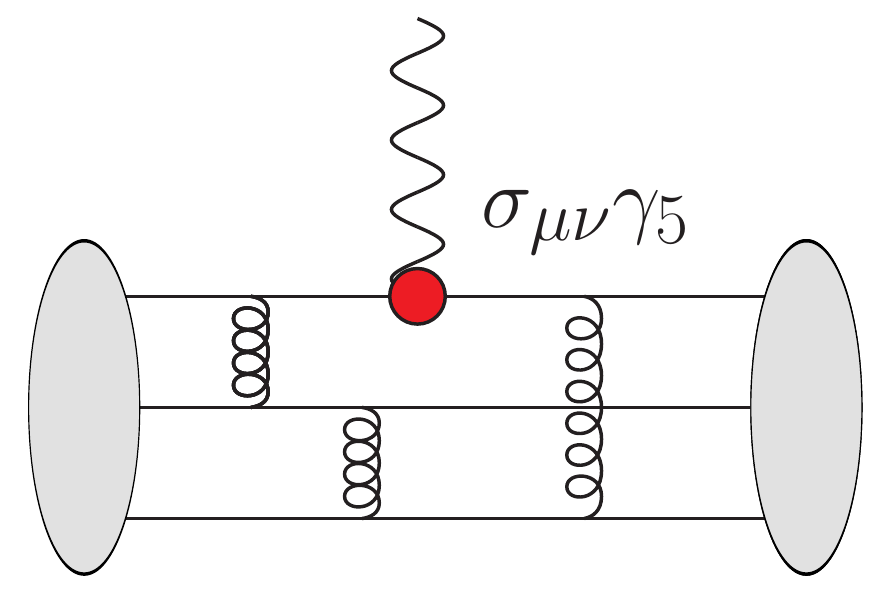}%
\hspace{0.025\textwidth}%
\includegraphics[width=0.2\textwidth]{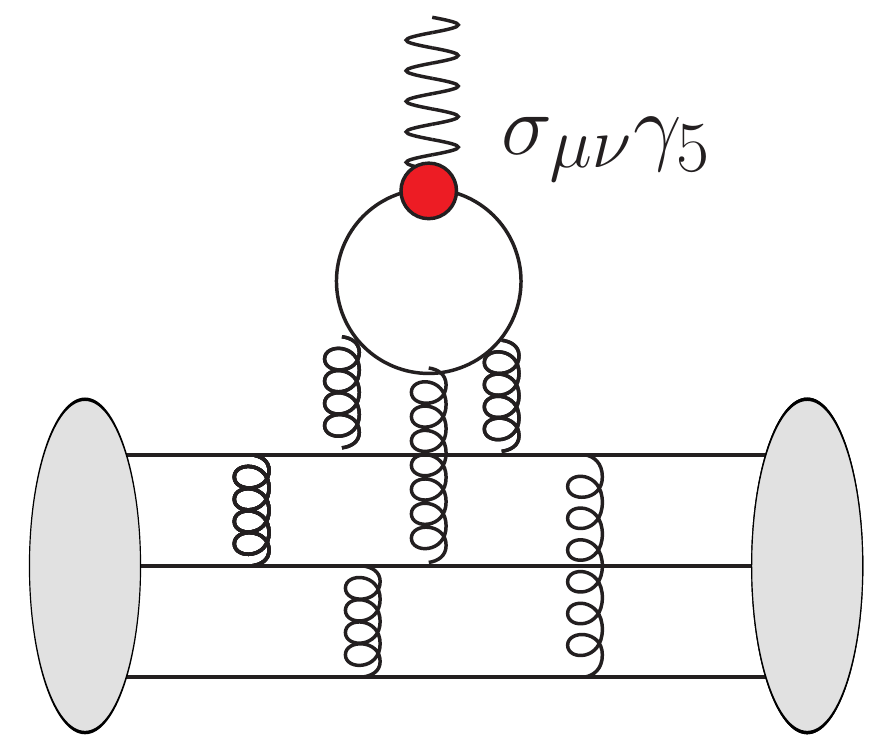}%
&%
\includegraphics[width=0.2\textwidth]{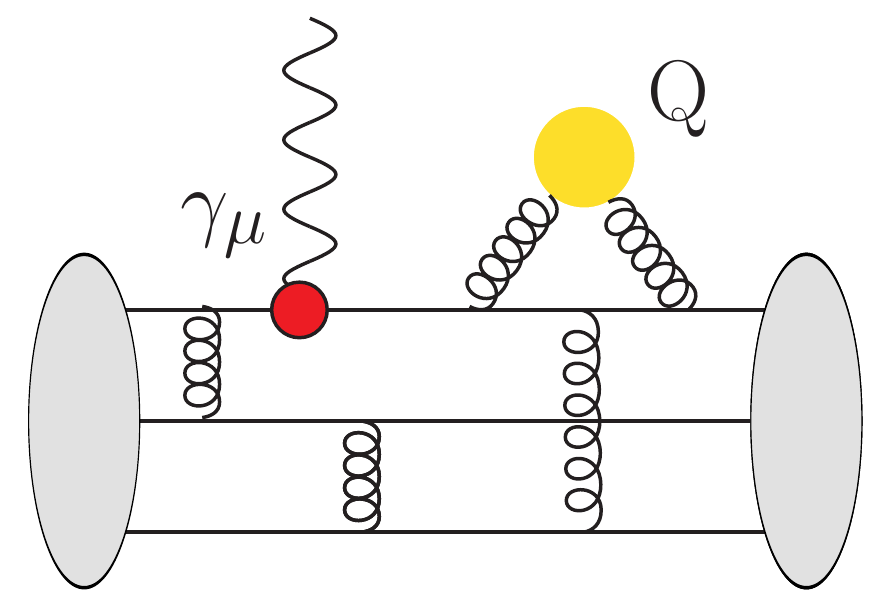}%
\hspace{0.025\textwidth}%
\includegraphics[width=0.2\textwidth]{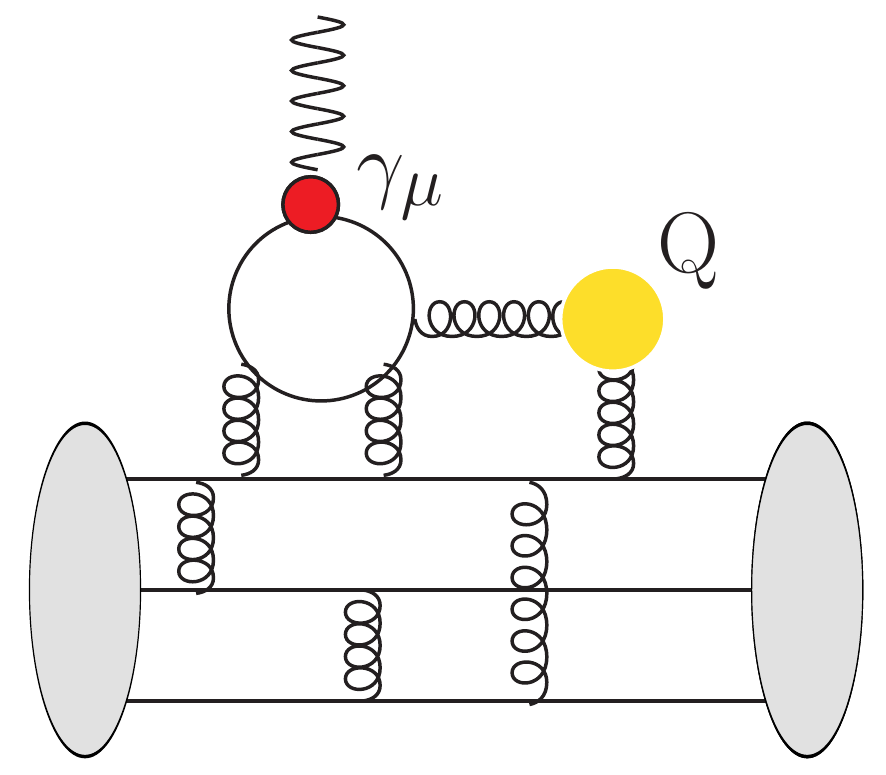}%
\end{tabular}
\caption{\it Calculating nEDM requires evaluating the interaction of
  the em current with the neutron. The contribution of connected and disconnected
  diagrams from the (left) qEDM and (right) $\Theta$-term.
\label{fig:calc}}
\vspace{-20pt}
\end{figure}
\end{center}

In the presence of \CPV, the electromagnetic current, defined as  $\delta {\cal L}/\delta A^\mu$, gets an 
additional term:
\begin{equation}
e \sum_q \overline{q} \gamma^{\mu} q \mathbin{{\longrightarrow}}
   e \sum_q \overline{q} \gamma^{\mu} q  + \epsilon^{\rho\sigma\nu\mu} p^\nu \sum_q d_q^\gamma \overline{q} \Sigma^{\mu\nu} q \,.
\label{eq:Vmu}
\end{equation}
The matrix elements of this leading qEDM (second) term are the flavor diagonal
tensor charges $g_T^q$: 
\begin{eqnarray}
  \left.\langle N \mid J^{\rm EM}_\mu \mid N \rangle\right|_{\not{\rm CP}}^{\rm qEDM,1}  &=& 
    \epsilon_{\mu\nu\kappa\lambda} q^\nu
    \left\langle N     \left| \left( 
       d_u^\gamma \bar u \Sigma^{\kappa\lambda} u + 
       d_d^\gamma \bar  d\Sigma^{\kappa\lambda} d +  
       d_s^\gamma \bar  s\Sigma^{\kappa\lambda} s + \ldots \right) \right| N \right\rangle \,, \nonumber \\
      & \lim_{q \to 0 } & d_u^\gamma g_T^u + d_d^\gamma g_T^d + d_s^\gamma g_T^s + \ldots \,.
\label{eq:quarkEDM}
\end{eqnarray}
The connected and disconnected (with $u$, $d$, $s$ and $c$ quark
loops) Feynman diagrams for the qEDM contribution are shown in
Fig.~\ref{fig:calc} (left).  
The first lattice results for the qEDM were given
in~\cite{Bhattacharya:2015wna} and phenomenological consequences for a
particular BSM theory, Split SUSY, were presented in
Ref.~\cite{Bhattacharya:2015esa}.  The key computational challenge to
evaluating Eq.~\eqref{eq:quarkEDM} is the calculation of the
disconnected contributions to $g_T^q$. These were shown to be small and 
noisy, and decrease with the quark
mass~\cite{Bhattacharya:2015wna}. The errors in the strange 
disconnected contribution were small enough 
to yield the continuum limit estimate 
$g_T^s = 0.008(9)$~\cite{Bhattacharya:2015wna}. In most
promising BSM theories, the new \CPV\ interactions are Yukawa like,
i.e., the couplings are proportional to the quark mass. Consequently,
the impact of the small value of $g_T^s$ with $O(1)$ error, gets magnified by the
quark mass ratio $2m_s/(m_u + m_d) = 27$.  Thus, it is important to 
improve the estimate of $g_T^s$ and, for the same reason,  calculate $g_T^c$.

%

Contributions from the CEDM and the next order in quark EDM arise due
to the change in the action, ${\cal L}_{\rm QCD} \longrightarrow {\cal
  L}_{\rm QCD}^{\cancel{\text{CP}}}$.  In an ideal world, the best way
to calculate these effects would be to simulate the lattice theory
using a discretized version of ${\cal L}_{\rm
  QCD}^{\cancel{\text{CP}}}$ and compute the matrix element
$\left.\langle N \mid J^{\rm EM}_\mu \mid N
\rangle\right|_{\cancel{\text{CP}}}$. This ideal approach is not
practical because \CPV\ interactions are complex and lattice
simulations of theories with a complex action are not yet realistic.

The method of choice is to treat these \CPV\ interactions as
perturbations in the small couplings $d_q^{\gamma}$ and $d_q^{G}$, and
expand the theory about ${\cal L}_{\rm QCD}$ (see Sec.~\ref{sec:reweight}). Then, the leading terms
that contribute to the nEDM are the matrix elements of the product of
the electromagnetic current $J^{\rm EM}_\mu$ and the 4-volume integral
of the  operators.  For the CEDM operator, the matrix
elements that have to be calculated are
\begin{eqnarray}
  \left.\langle N \mid J^{\rm EM}_\mu \mid N \rangle\right|_{\cancel{\text{CP}}}^{\rm CEDM} &=& 
       \left\langle N \left| J^{\rm EM}_\mu \int d^4 x [
       \left( d_u^G \bar u \Sigma_{\nu\kappa} u + 
              d_d^G \bar d \Sigma_{\nu\kappa} d + 
              d_s^G \bar s \Sigma_{\nu\kappa} s + \ldots  \right) \tilde G^{\nu\kappa} ] \right| N \right\rangle  \,, 
\label{eq:chromoEDM}
\end{eqnarray}
as discussed in Sec.~\ref{sec:reweight}. Similarly for the next order qEDM 
\begin{eqnarray}
       \left.\langle N \mid J^{\rm EM}_\mu \mid N \rangle\right|_{\cancel{\text{CP}}}^{\rm qEDM,2} &=& 
       \left\langle N \left| J^{\rm EM}_\mu \int d^4 x [
       \left( d_u^F \bar u \Sigma_{\nu\kappa} u + 
              d_d^F \bar d \Sigma_{\nu\kappa} d + 
              d_s^F \bar s \Sigma_{\nu\kappa} s + \ldots  \right) \tilde F^{\nu\kappa} ]\right| N \right\rangle  \,.
\label{eq:quarkEDM2}
\end{eqnarray}
This second qEDM term has two electromagnetic interactions and is therefore 
expected to be smaller than the leading term given in Eq.~\eqref{eq:quarkEDM}. For this reason 
we neglect its contribution. 
Finally, the contribution of the $\Theta$-term requires calculating the matrix element: 
\begin{equation}
  \left.\langle N \mid J^{\rm EM}_\mu \mid N \rangle\right|_{\not{\rm CP}}^{\Theta} =
       \left\langle \left| J^{\rm EM}_\mu \ \int d^4 x\ \Theta  G^{\mu\nu}
       \tilde G^{\mu\nu} \right| n \right\rangle \,.
\label{eq:ThetaEDM}
\end{equation}
This correlation of $J_\mu$ with topological charge $\cal Q$ gives rise to 
the 2 diagrams shown in Fig.~\ref{fig:calc} (right).

\section{Simulating QCD Including Complex \CPV\ CEDM and $\gamma_5$ operators}
\label{sec:reweight}

%
\begin{figure}[tp]
\begin{eqnarray*}
\vcenter{\hbox{\includegraphics[width=0.10\textwidth]{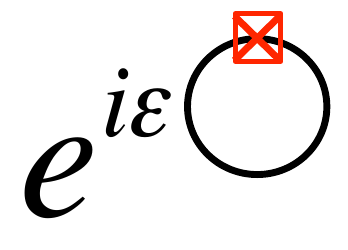}  }}
\left(\;\vcenter{\hbox{\includegraphics[width=0.38\textwidth]{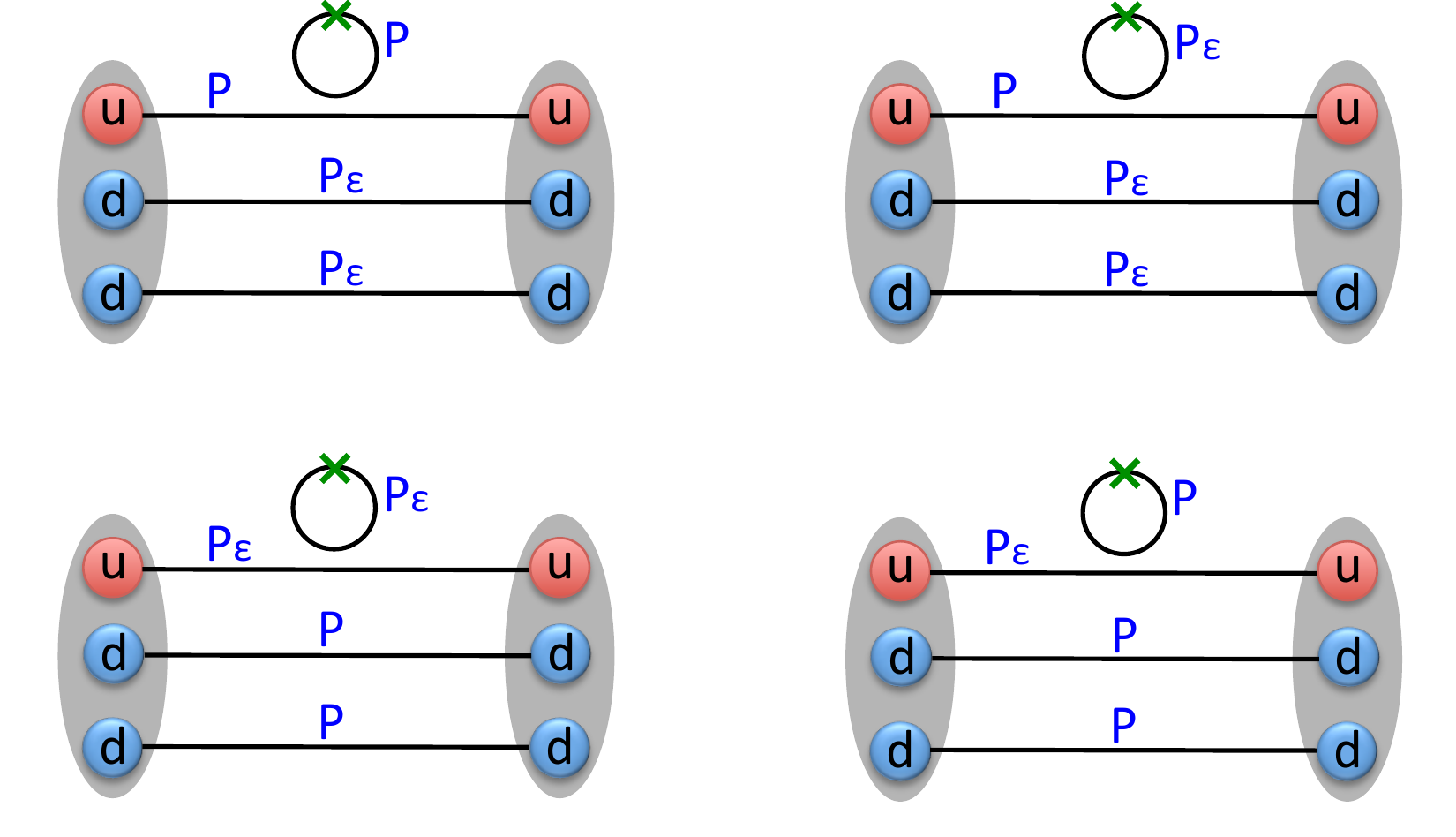}}} + 
 \vcenter{\hbox{\includegraphics[width=0.38\textwidth]{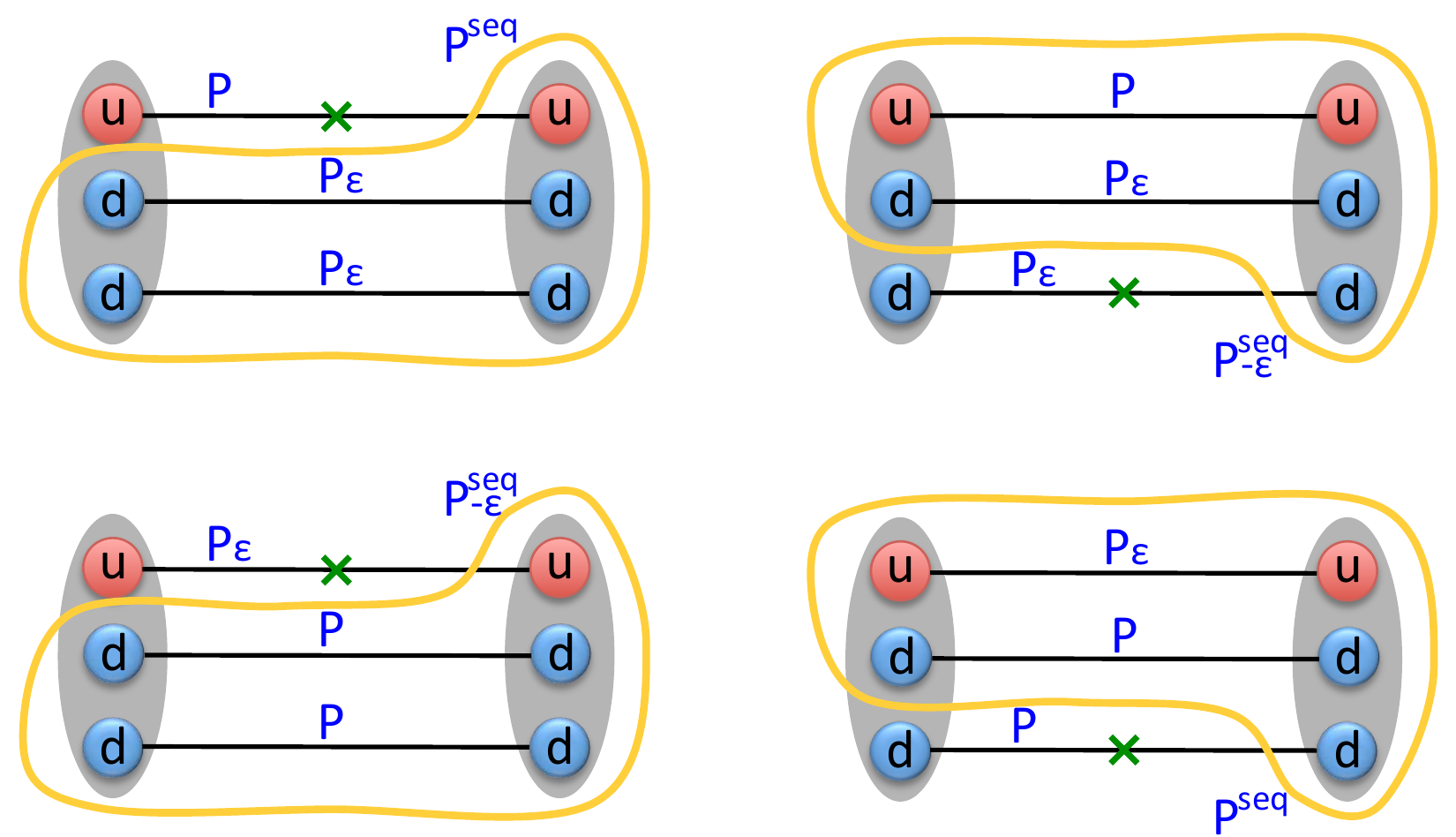}}}\;\right)
\end{eqnarray*}
\caption{CEDM contribution is the reweighting factor times
  the sum of connected and disconnected diagrams. }
\label{fig:cEDMfull}
\end{figure}

The QCD Lagrangian becomes complex with the addition of any CP
violating interaction. A robust cost-effective method for simulating
theories with complex actions does not exist.  To calculate the
contribution of the \CPV\ interactions to the nEDM, we work within the
framework of the Schwinger source method.  We start by noting
that the fermion part of the Lagrangian, ${\cal L}_{\rm
  QCD}^{\cancel{\text{CP}}}$ in Eq.~\eqref{eq:Lcpv}, involves only
quark bilinear operators in presence of the qEDM and CEDM interactions. 
The fermion fields in the path integral can, therefore, still be
integrated out.  There are, however, two modifications to the standard
calculations of 3-point functions due to the presence of
\CPV\ interactions treated as sources. First, the Dirac operator, in the presence of say the
CEDM term, becomes
\begin{equation}
   \left.\Dslash + m - \frac r2 D^2 + c_{sw} \Sigma^{\mu\nu} G_{\mu\nu}\right. \mathbin{{\longrightarrow}}
   \left.\Dslash + m - \frac r2 D^2 + \Sigma^{\mu\nu} ( c_{sw} G_{\mu\nu} + i \epsilon {\tilde G_{\mu\nu}} )\right.  \,.
\label{eq:Dcpv}
\end{equation}
Quark propagators calculated with this modified Dirac operator include the
insertion of the CEDM operator at all space-time points. Second, the Boltzmann weight of all
gauge configurations generated without the CEDM term in the action need to be
modified by reweighting them by the ratio of the
determinants of the Dirac operators for the two theories--with and without \CPV\ terms:
\begin{eqnarray}
\frac {\det ( \Dslash + m - \frac r2 D^2 + \Sigma^{\mu\nu} ( c_{sw} G_{\mu\nu} + i \epsilon {\tilde G_{\mu\nu}} )}
          {\det ( \Dslash + m - \frac r2 D^2 + c_{sw} \Sigma^{\mu\nu} G_{\mu\nu} )}\span\omit\span\hfill\nonumber\\
&=&
\exp \Tr \ln \left[1 + i \epsilon\, {\Sigma^{\mu\nu} \tilde G_{\mu\nu}} ( \Dslash + m - \frac r2 D^2 + c_{sw} \Sigma^{\mu\nu} G_{\mu\nu} )^{-1}\right]
\\
&\approx&
\exp \left[ i \epsilon \Tr {\Sigma^{\mu\nu} \tilde G_{\mu\nu}} ( \Dslash + m - \frac r2 D^2 + c_{sw} \Sigma^{\mu\nu} G_{\mu\nu} )^{-1}\right]\,.
\label{eq:reweight}
\end{eqnarray}%
Since the coupling $\epsilon$ for the CEDM term for all quark flavors
is small, we expect the leading order reweighting factor to be a good
approximation.  This `reweighting factor' for each gauge configuration
is, to leading order, the integral over all spacetime points of the
value of a closed quark loop with the insertion of the CEDM
operator. 

There is an additional challenge: The CEDM operator has an UV divergent
mixing with the ``$\gamma_5$-operator, $\overline{q} \gamma_5
q/a^2$~\cite{Bhattacharya:2015rsa}. Thus, one has to (i) calculate the
matrix element of the ``$\gamma_5$-operator'', which we do in the same
way as for the CEDM operator, (ii) calculate a similar reweighting
factor, and (iii) handle the $1/a^2$ divergent mixing.  Since the
reweighting is an overall phase, and configuration to configuration
fluctuations can be large, it is therefore important to demonstrate
the quality of the signal in $F_3(0)$ after reweighting for both the
CEDM and $\gamma_5$ terms.  In Sec.~\ref{sec:Num}, we describe the
progress we have made towards addressing these challenges. There is 
also a mixing of the CEDM operator with the $G^{\mu\nu} \tilde G^{\mu\nu}$ 
operator. Addressing this mixing is part of the calculation of the 
$\Theta-$term per Eq.~\eqref{eq:ThetaEDM}~\cite{Bhattacharya:2015rsa}.

The Schwinger source method has allowed us to recast
the challenging calculation of 4-point functions given in
Eq.~\eqref{eq:chromoEDM} to the difficult calculation of 3-point
functions~\cite{Bhattacharya:2016oqm}.  The Feynman diagrams needed to
calculate $\langle N | J_\mu^{\rm EM}(q) | N \rangle $ in the presence
of CEDM and $\gamma_5$ interactions are illustrated in
Fig.~\ref{fig:cEDMfull}.  Starting with an ensemble of gauge
configurations generated with a standard lattice action, for example
the Wilson-clover action, the steps in the calculation are:
\begin{itemize}
\item
Calculate propagators, labeled $P$ in Fig.~\ref{fig:cEDMfull}, using
the standard Wilson-clover Dirac operator. We assume isospin symmetry
so the propagators for $u$ and $d$ quarks are numerically the same.
\item
Calculate a second set of propagators with the CEDM term included 
with coefficient $\epsilon$ in the Wilson-clover matrix: 
{\begin{equation}
   \left(\Dslash + m - \frac r2 D^2 + c_{sw} \Sigma^{\mu\nu} G_{\mu\nu}\right)^{-1} \mathbin{{\longrightarrow}}
   \left(\Dslash + m - \frac r2 D^2 + \Sigma^{\mu\nu} ( c_{sw} G_{\mu\nu} + i \epsilon {\tilde G_{\mu\nu}} )\right)^{-1}\,.
\end{equation}}%
These propagators, labeled $P_\epsilon$, include the full effect of
inserting the CEDM operator at all possible points along
them. The cost of inversion increases by about $7\%$ with respect to
$P$, however, using $P$ as the starting guess in the inversion for
$P_\epsilon$ reduces the number of iterations required by 20--40\%
depending on the quark mass. The overall average cost of $P_\epsilon$
is found to be about 80\% of $P$.
\item
Using $P$ and $P_\epsilon$, construct 4 kinds of sequential
sources, labeled $P_u^{\rm seq}$, $P_d^{\rm seq}$,
$P_{-\epsilon,u}^{\rm seq}$, and $P_{-\epsilon,d}^{\rm seq}$, at the
sink time-slice corresponding to the insertion of a neutron at zero
momentum. The subscripts $u$ or $d$ in $P_u^{\rm seq}$ and $P_d^{\rm
  seq}$, and similarly in $P_{-\epsilon,u}^{\rm seq}$ and
$P_{-\epsilon,d}^{\rm seq}$, denote the flavor of the uncontracted spinor in the
neutron source. The minus sign in the coupling, $-\epsilon$, accounts for 
the backward moving propagator. 
\item
If using the coherent sequential source method, then $N$ sources at different time-slices 
for the $N$ different calculations being done simultaneously on that configuration, are
added together. Sequential propagators with these (coherent)
sequential sources are calculated by inverting the Dirac operator with
and without the CEDM term as appropriate. The 4 types of (coherent)
sequential propagators are shown by the block of 4 correlation functions
in the right half of Fig.~\ref{fig:cEDMfull}.
\item
Using the two original and the four sequential propagators, calculate
the connected 3-point function. These contractions give the
four 3-point functions shown on the right in Fig.~\ref{fig:cEDMfull}. 
Note that a separate insertion of $J_\mu^{\rm EM}$ is done 
on the $u$ and $d$ quark lines. 
\item
Calculate the disconnected quark loop with the insertion of electromagnetic 
current at zero momentum for each of the quark flavors, $u,\ d, \, s,\ c$ (and $b$ 
if needed) using the Dirac propagator with and without the CEDM term. 
\item
Calculate the correlation of these quark loops with the appropriate
nucleon 2-point correlation functions. The 4 types of disconnected
Feynman diagrams required are illustrated in the left of
Fig.~\ref{fig:cEDMfull}.
\item
To correct for the omission of the chromo EDM operator in the action
during the generation of the gauge configurations, calculate the volume
integral of the quark loop with the CEDM insertion for each
flavor. Multiply the sum of the connected and disconnected
contributions by the reweighting factor--exponential of the estimate
of the ``CEDM loop'' for the configuration with the appropriate
coupling factor $i\epsilon$. This is shown by the overall factor in
Fig.~\ref{fig:cEDMfull}.
\item
Repeat the calculation for different values of $\epsilon$ that bracket
the expected numerical values of the couplings $d_q^G$ for the various
quark flavors.
\item
Repeat the whole calculation for the $\gamma_5$-operator instead of the CEDM operator. 
\end{itemize}

\section{Matrix Elements and Form Factors from 3-point Functions}
\label{sec:3pt}

\begin{figure}[tbhp]
\begin{center}
\vspace*{-0.2in}
{
\includegraphics[width=0.46\textwidth]{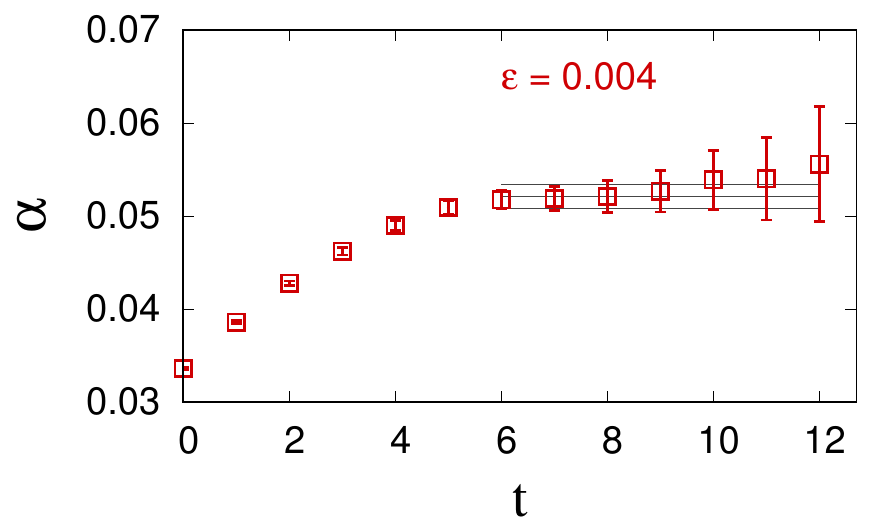}%
\includegraphics[width=0.46\textwidth]{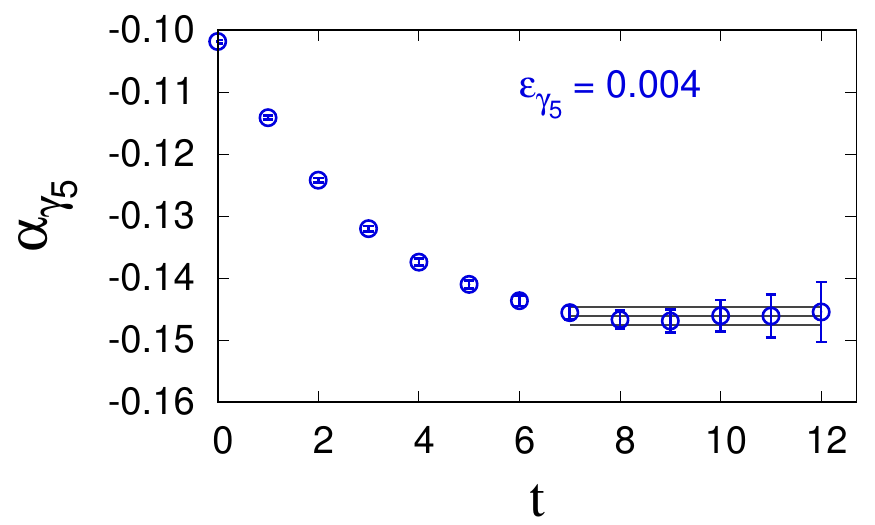}%
}
{
\includegraphics[width=0.46\textwidth]{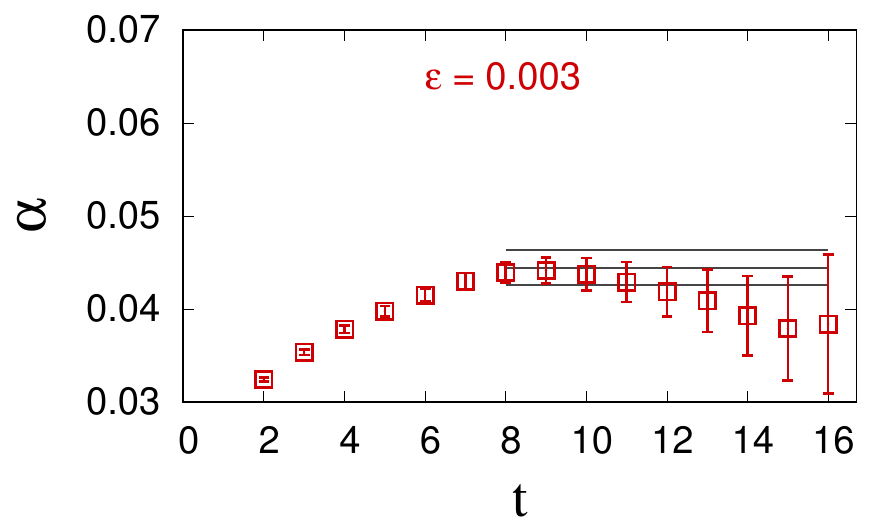}%
\includegraphics[width=0.46\textwidth]{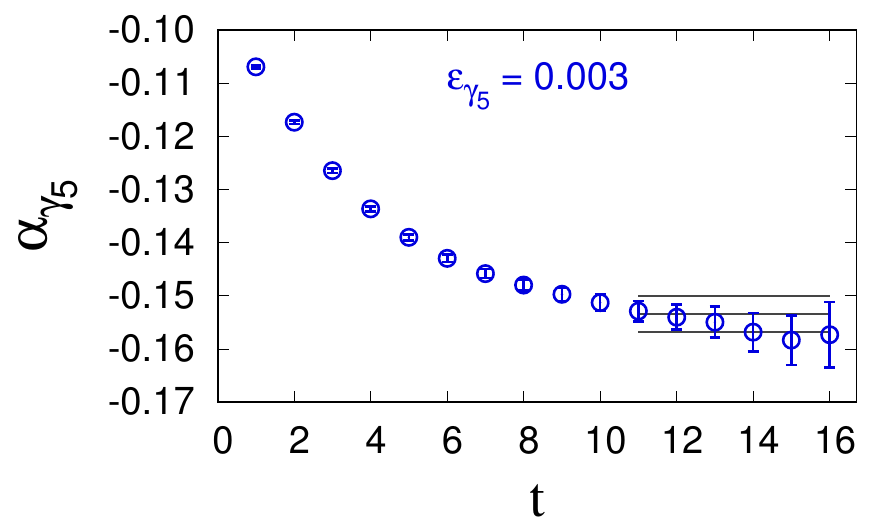}%
}
\end{center}
\vspace{-18pt}
\caption{\it The phase $\alpha $ extracted from the plateau in the imaginary part
  of the neutron 2-point function with $\epsilon=0.004$ on the $a12m310$ ensemble (top) and 
  $\epsilon=0.003$ on the $a09m310$ ensemble (bottom).  The plots on the left are for the CEDM 
  operator and on the right for the $\gamma_5$ mixing operator. 
}
\label{fig:alpha}
\end{figure}

\begin{figure}[tbhp]
\begin{center}
{
\vspace*{-0.8in}
\includegraphics[width=0.4\textwidth]{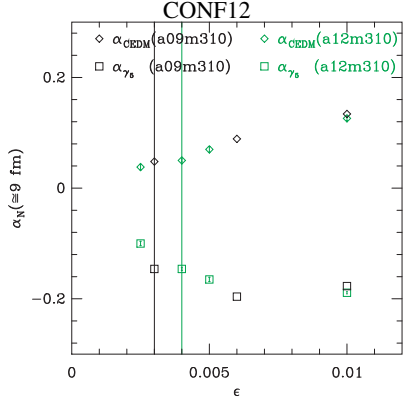}%
}
\vspace*{-0.6in}
\end{center}
\caption{\it Linear behavior of the phase $\alpha $ versus $\epsilon$ 
            with the insertion of the CEDM and the $\gamma_5$ operators. }
\label{fig:alpha-lin}
\end{figure}

The matrix elements, defined in
Eqs.~\eqref{eq:quarkEDM},~\eqref{eq:chromoEDM}
and~\eqref{eq:ThetaEDM}, are extracted from the vacuum to vacuum
3-point correlation function: insertion of the electromagnetic current
at times $t$ between the neutron source and sink operators at time $0$
and $T$:
\begin{equation}
\langle \Omega | {N(\vec0,0) J^{\rm EM}_\mu(\vec q,t) N^\dagger(\vec p,T)} | \Omega \rangle =
 \sum_{N,N'} u_N e^{-M_N t}\; {\langle N | J^{\rm EM}_\mu(q) | N' \rangle}\; e^{-E_{N'} (T-t)}{\overline u}_N'\,.
\label{eq:3pt}
\end{equation}%
where in the right hand side we have used the expansion in a complete
basis of states \(|N\rangle, |N'\rangle, \ldots\) that couple to the
neutron interpolating operator $N$.  We use $ N \equiv \epsilon^{abc}
[{d}^{a{\rm T}} C \gamma_5 (1+\gamma_4) u^b] d^c $ where $C=\gamma_0
\gamma_2$ is the charge conjugation matrix, $a,\ b, c$ are the color
indices, $u,\ d$ are the quark flavors, \(u_N\) is the free neutron
spinor and \(M_N\) is the neutron mass.  In the
presence of \CPV, the nucleon 2-point function is 
\begin{equation}
\langle \Omega | {N(\vec0,0) N^\dagger(\vec p,t)} | \Omega \rangle|_{\lim{t\to \infty}} = 
A^2 u_N \, e^{-p_4 t} \, {\overline u}_N =  A^2 e^{-p_4 t} {e^{i\alpha_N \gamma_5}} (i\slashed p + m_N) {e^{i\alpha_N \gamma_5}}\,,
\label{eq:phase}
\end{equation}
where the phase angle \(\alpha_N\) arises due to the \CPV\ coupling
$\epsilon$ and depends on it. To determine the desired region of linearity of $\alpha$ versus
$\epsilon$ we show the imaginary part of the neutron
2-point function in Fig.~\ref{fig:alpha}. It demonstrates the quality of the
signal for $\alpha$ for appropriately small values of $\epsilon$ on two different
ensembles. Fig.~\ref{fig:alpha-lin} shows the expected linear 
behavior of $\alpha$ versus $\epsilon$ over a range of small $\epsilon$. 

Once the matrix element of the electromagnetic current, $J^{\rm EM}_\mu(q)$ defined
in Eq.~\ref{eq:Vmu}, within the nucleon state is extracted using Eq.~\eqref{eq:3pt}, it is parameterized in
terms of four Lorentz covariant form factors $F_1$, $F_2$, $F_A$ and
$F_3$:
\begin{eqnarray}
\langle N | J_\mu^{\rm EM}(q) | N \rangle & = &
   \overline {u}_N \left[ 
         \gamma_\mu\;F_1(q^2) \quad + \quad i \sigma^{\mu\nu}\; q_\nu\; \frac{F_2(q^2)}{2 M_N} 
         \right. \nonumber\\[1\jot]
&&\left. + \quad  {(2 i\,M_N \gamma_5 q_\mu - \gamma_\mu \gamma_5 q^2)\;\frac{F_A(q^2)}{M_N^2}} \quad + \quad
         \sigma^{\mu\nu}\; q_\nu \gamma_5\;\frac{F_3(q^2)}{2 M_N} \right] u_N\,,
\label{eq:FFdef}
\end{eqnarray}
\(F_1\) and \(F_2\) are the standard Dirac and Pauli form
factors.  The anapole form factor \(F_A\) violates parity P and 
the electric dipole form factor \(F_3\) violates P and CP.
$F_3$ is extracted from the different matrix elements by using different combinations of 
the momentum transfer $q_\mu$ and spin projections. 
The zero momentum limit of these form
factors gives the charges and dipole moments: the
electric charge is \(F_1(0) = 1\) and the anomalous magnetic
dipole moment is \(F_2(0)/2 M_N \). 
The contribution of the matrix element of each \CPV\ interaction defined in
Eqs.~\eqref{eq:quarkEDM},~\eqref{eq:chromoEDM},
and~\eqref{eq:ThetaEDM} to the electric dipole moment of the neutron is given by 
$
            d_n = \lim_{q^2\to0} {F_3(q^2)}/{2 M_n} \,.
$

\section{Status of Numerical Calculations}
\label{sec:Num}

So far, we have performed numerical calculations on two 2+1+1-flavor HISQ
ensembles generated by the MILC
collaboration~\cite{Bazavov:2012xda}. The first, labeled $a12m310$,
has $a=0.12$~fm and $M_\pi=310$~MeV and the second, labeled $a09m310$,
has $a=0.09$~fm and $M_\pi=310$~MeV. The correlation functions are
constructed using Wilson-clover fermions.  On the $a12m310$ ($a09m310$)
ensemble we have analyzed 400 (270) configurations with 64
measurements on each configuration.  The ensembles used and our
strategy for the calculation of 2- and 3-point functions with this
clover-on-HISQ approach are described in
Ref.~\cite{Bhattacharya:2016zcn}.

\begin{figure}[htbp]
  \centering 
{
  \includegraphics[width=0.24\textwidth]{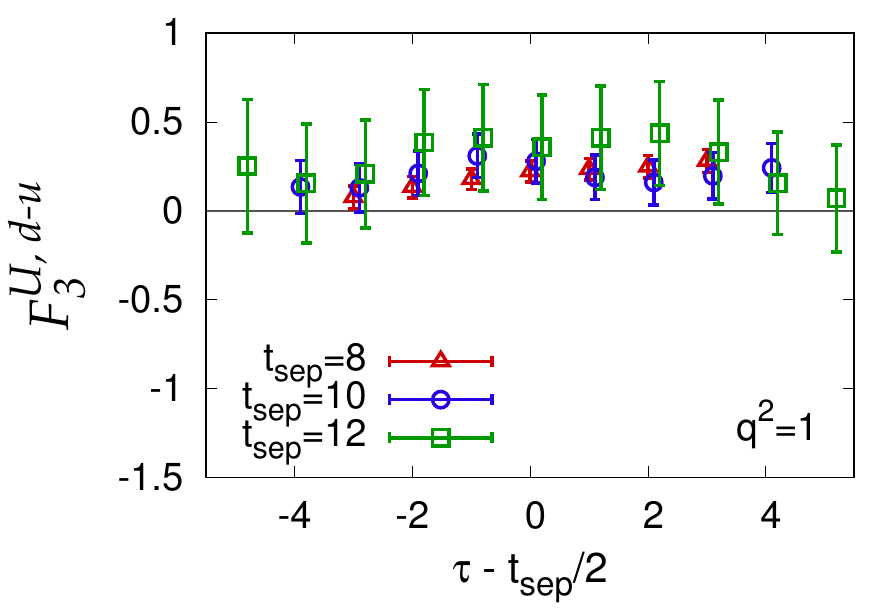}
  \includegraphics[width=0.24\textwidth]{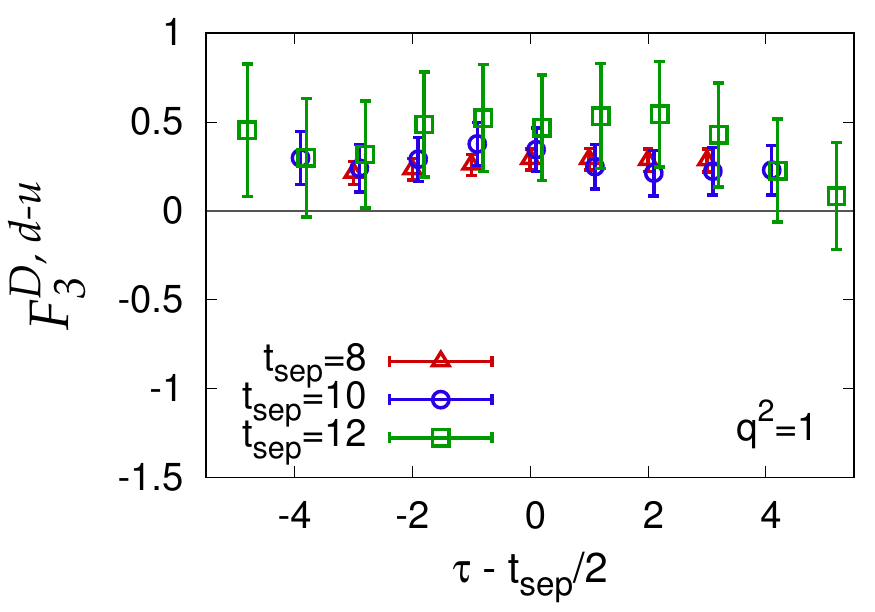}
  \includegraphics[width=0.24\textwidth]{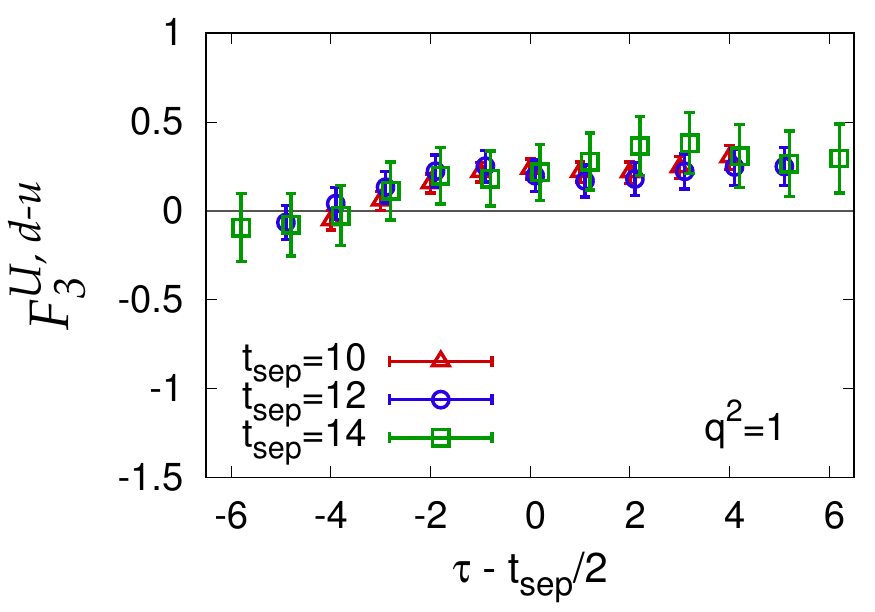}
  \includegraphics[width=0.24\textwidth]{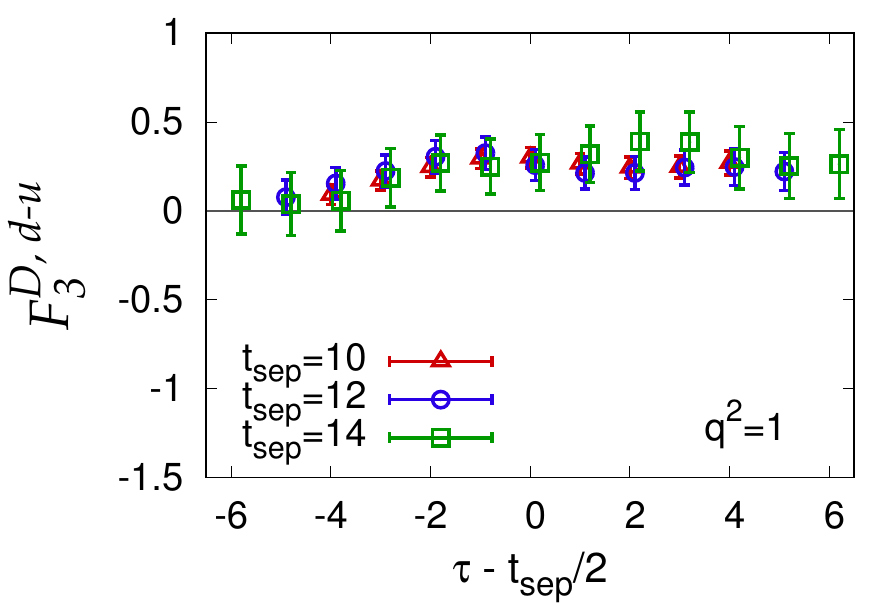}
}
{
  \includegraphics[width=0.24\textwidth]{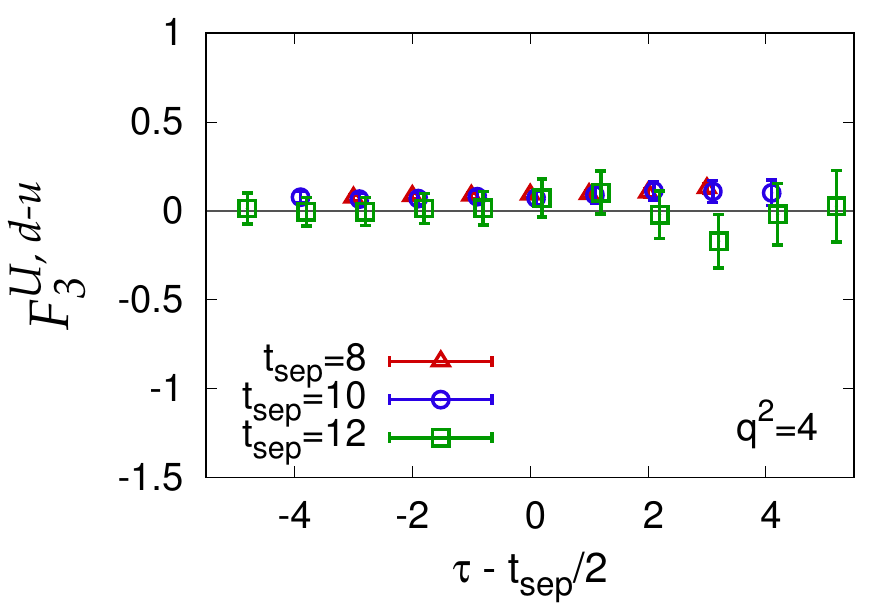}
  \includegraphics[width=0.24\textwidth]{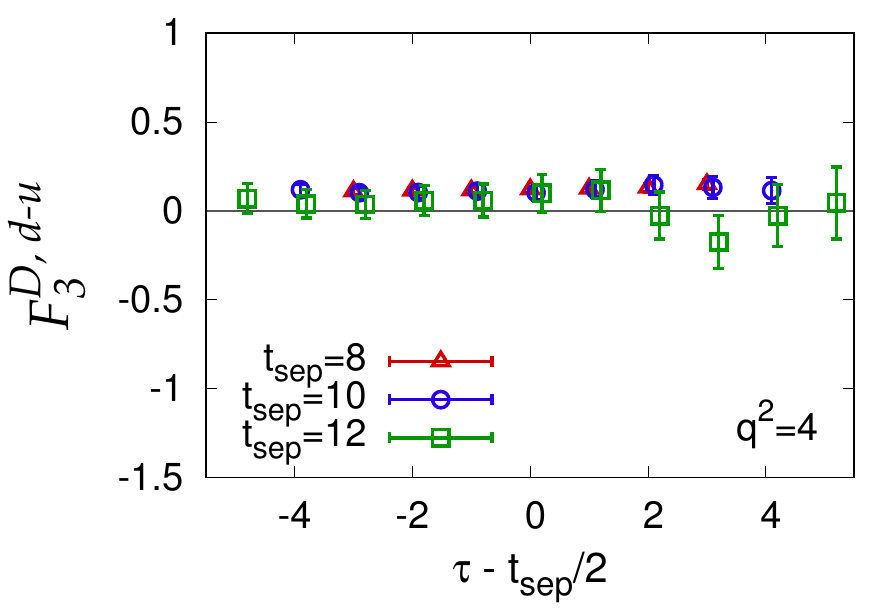}
  \includegraphics[width=0.24\textwidth]{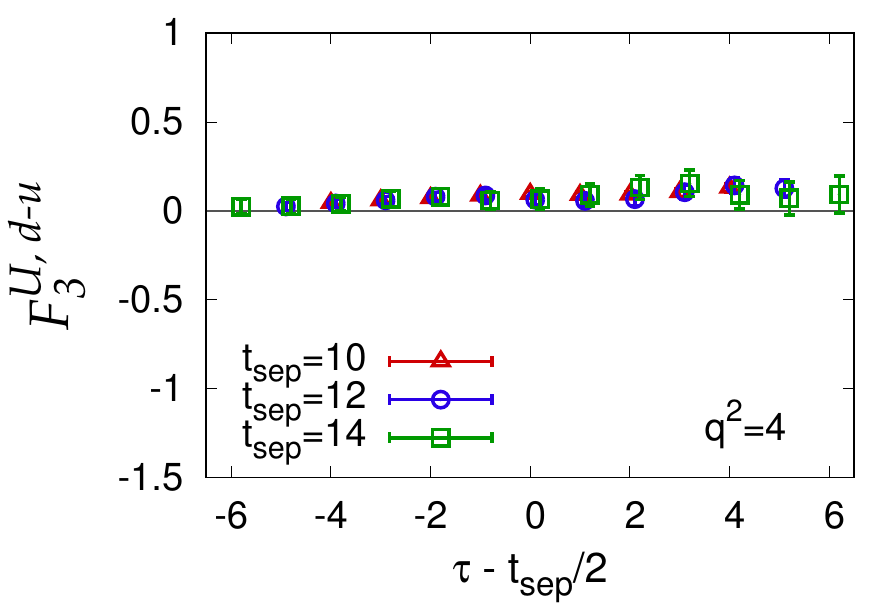}
  \includegraphics[width=0.24\textwidth]{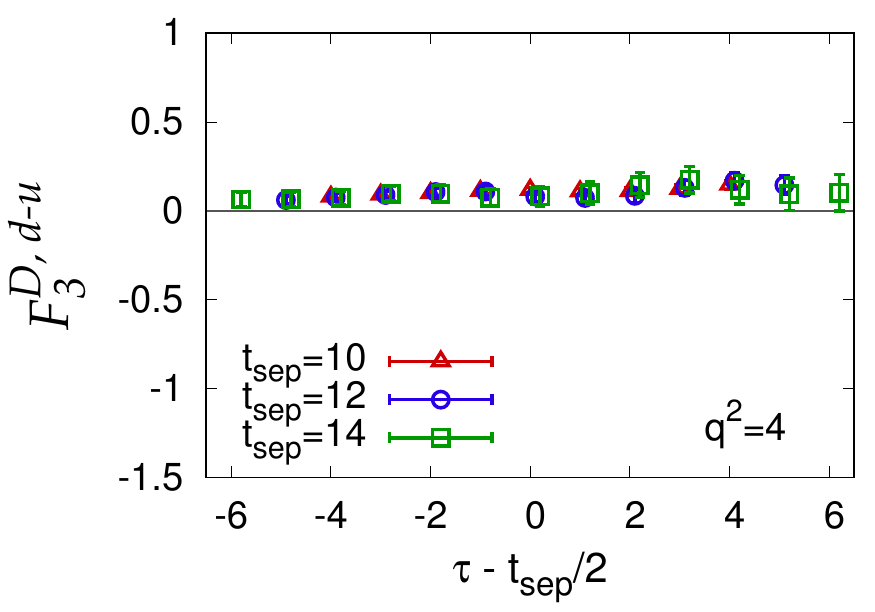}
}
  \caption{Illustration of the signal in $F_3$ with the inclusion of
    the CEDM term.  The data in the top row are for $\vec{p} = (1,0,0)
    \times 2\pi/La$ in the following order: insertion on $u$ and $d$
    quarks for the $a12m310$ ensemble followed by insertion on $u$ and
    $d$ quarks for the $a09m310$ ensemble. The plots in the second row
    are in the same order except with $\vec{p} = (2,0,0) \times
    2\pi/La$.  }
\label{fig:F3CEDM}
\end{figure}

\begin{figure}[htbp]
  \centering 
{
  \includegraphics[width=0.24\textwidth]{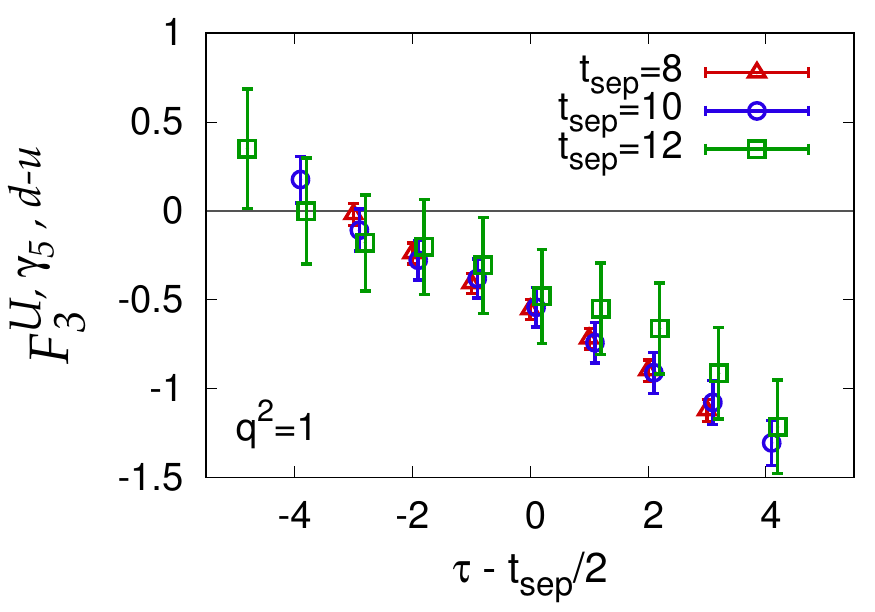}
  \includegraphics[width=0.24\textwidth]{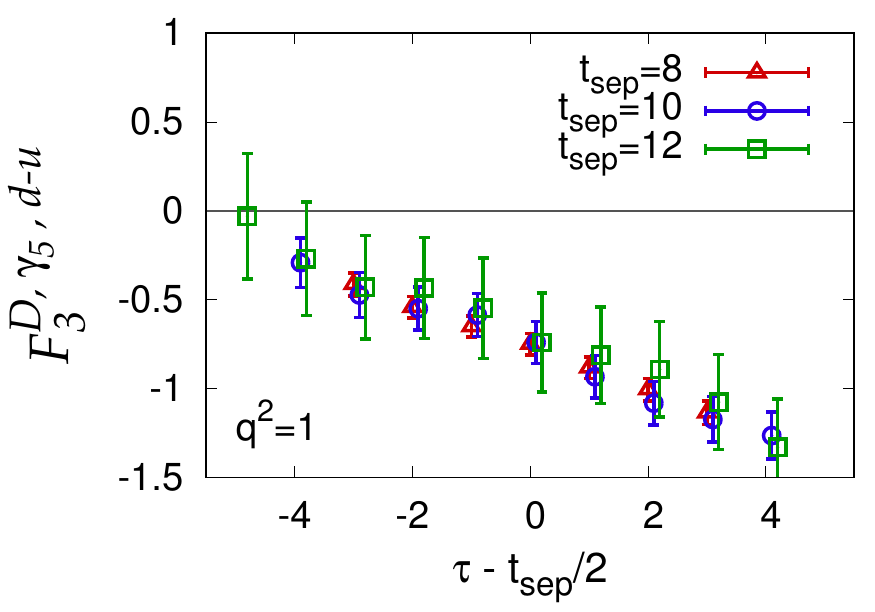}
  \includegraphics[width=0.24\textwidth]{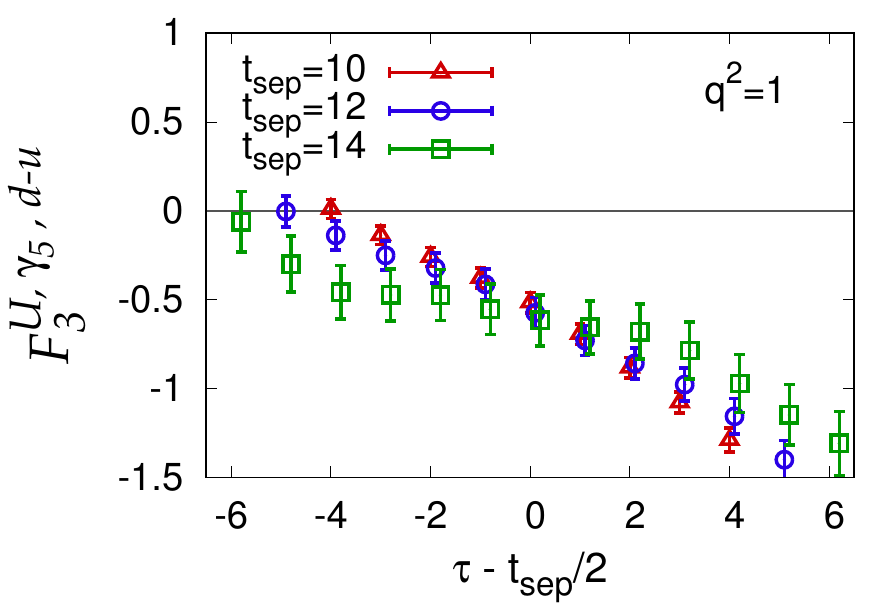}
  \includegraphics[width=0.24\textwidth]{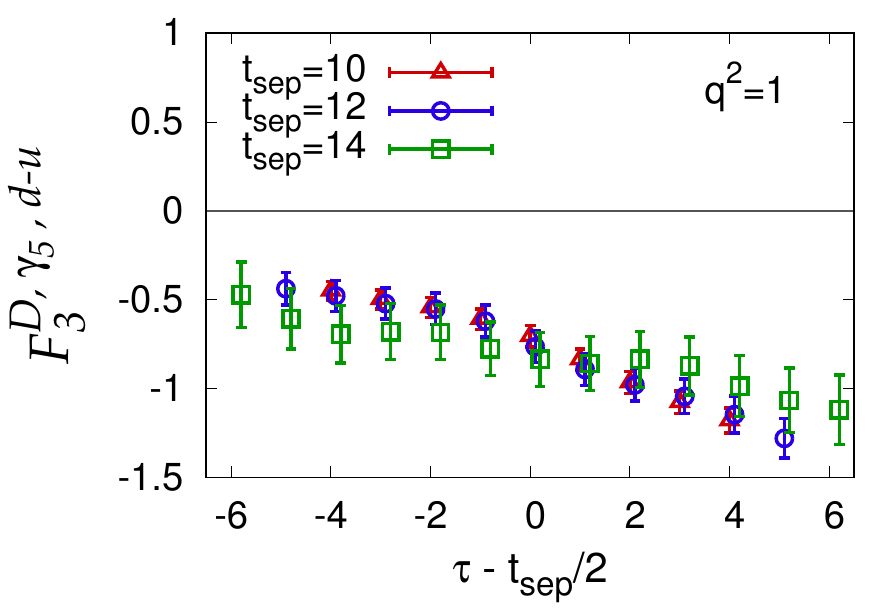}
}
{
  \includegraphics[width=0.24\textwidth]{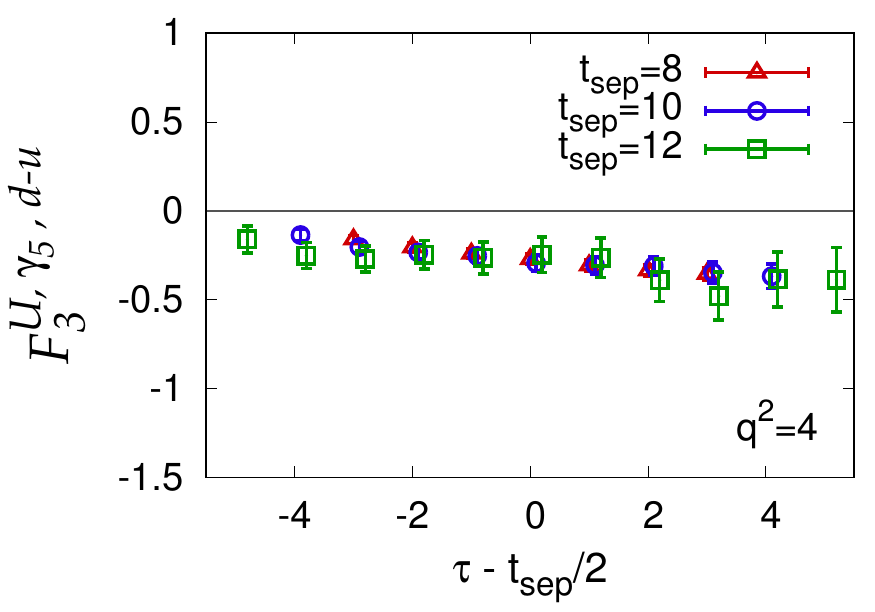}
  \includegraphics[width=0.24\textwidth]{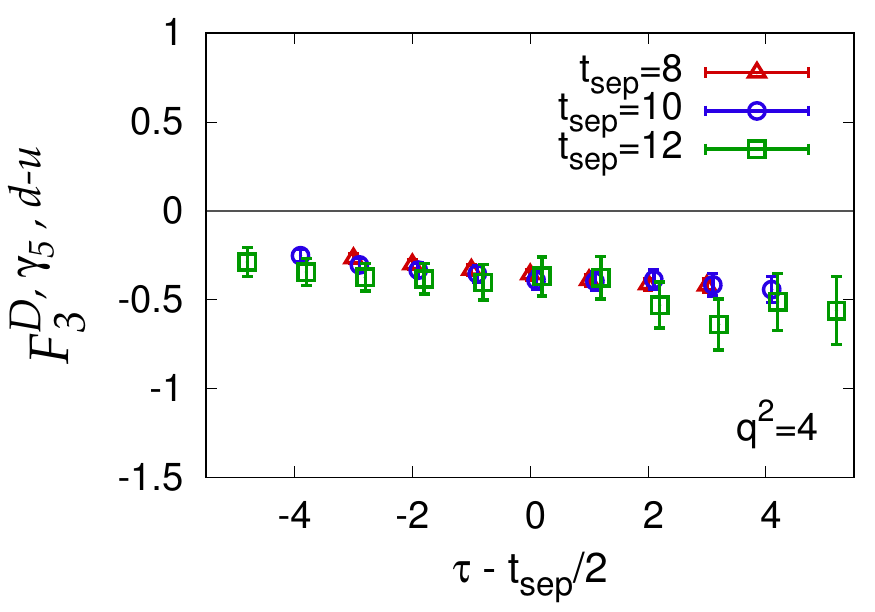}
  \includegraphics[width=0.24\textwidth]{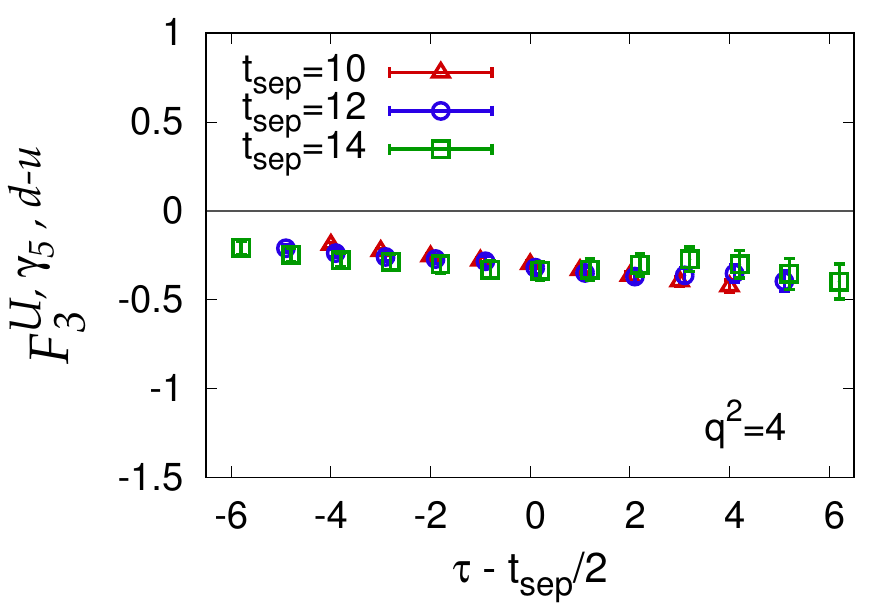}
  \includegraphics[width=0.24\textwidth]{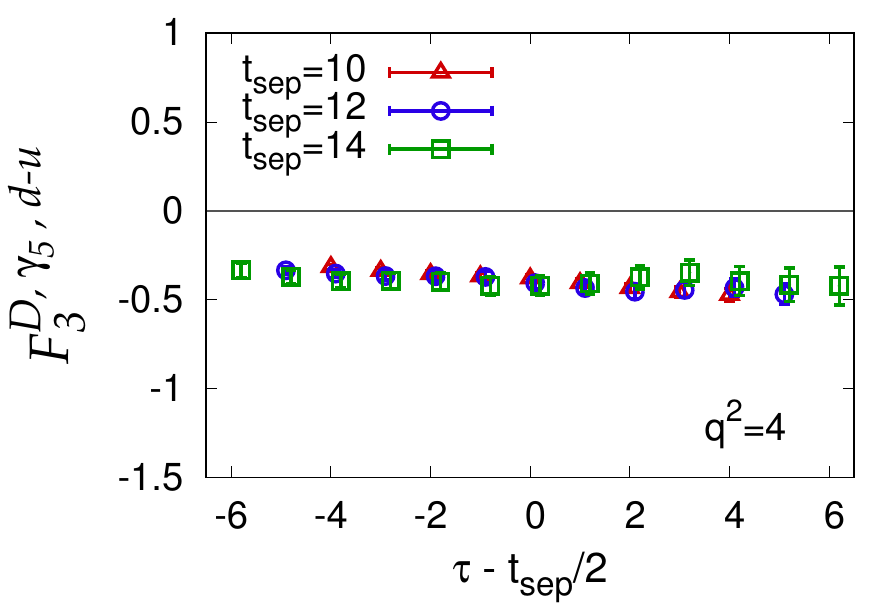}
}
  \caption{Illustration of the signal in $F_3$ with the inclusion of
    the $\gamma_5$ term.  The data in the top row are for $\vec{p} =
    (1,0,0) \times 2\pi/La$ in the following order: insertion on $u$
    and $d$ quarks for the $a12m310$ ensemble followed by insertion on
    $u$ and $d$ quarks for the $a09m310$ ensemble. The plots in the
    second row are in the same order except with $\vec{p} = (2,0,0)$.
  }
\label{fig:F3g5}
\end{figure}

In Fig.~\ref{fig:F3CEDM}, we show the data for $F_3$ from the
connected diagrams in the presence of the CEDM term, and in
Fig.~\ref{fig:F3g5}, the data for $F_3$ in the presence of the
$\gamma_5$ term.  The data are presented for three values of the
source-sink separation $\tsep=8, 10$ and $12$. The excited-state
contamination in the matrix elements, and thus in $F_3$, is removed by
taking the $\tsepi$ limit. These figures show that an acceptable
signal-to-noise ratio can be obtained to estimate $F_3$ with $O(1)$
errors, our first goal for the CEDM calculations.

\begin{figure}[htbp]
  \centering 
{
  \includegraphics[width=0.24\textwidth]{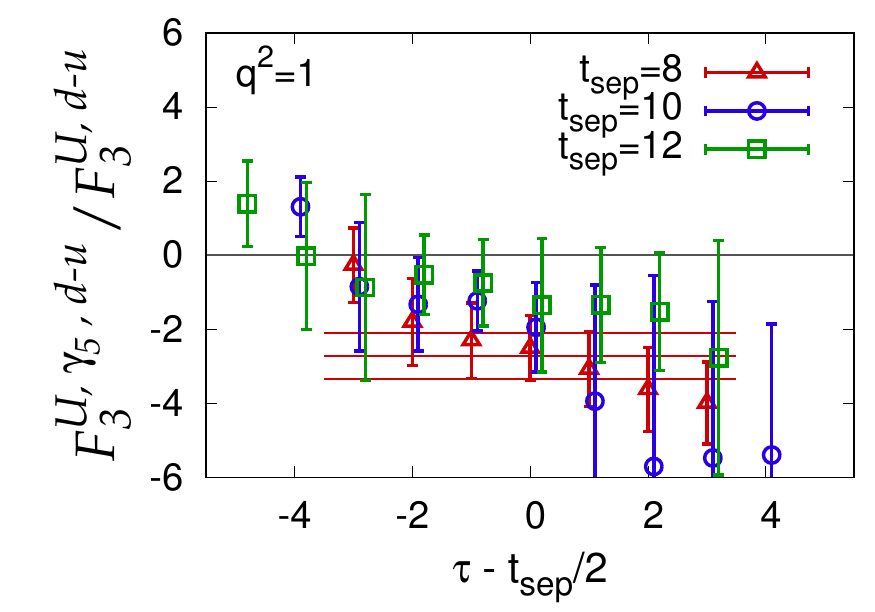}
  \includegraphics[width=0.24\textwidth]{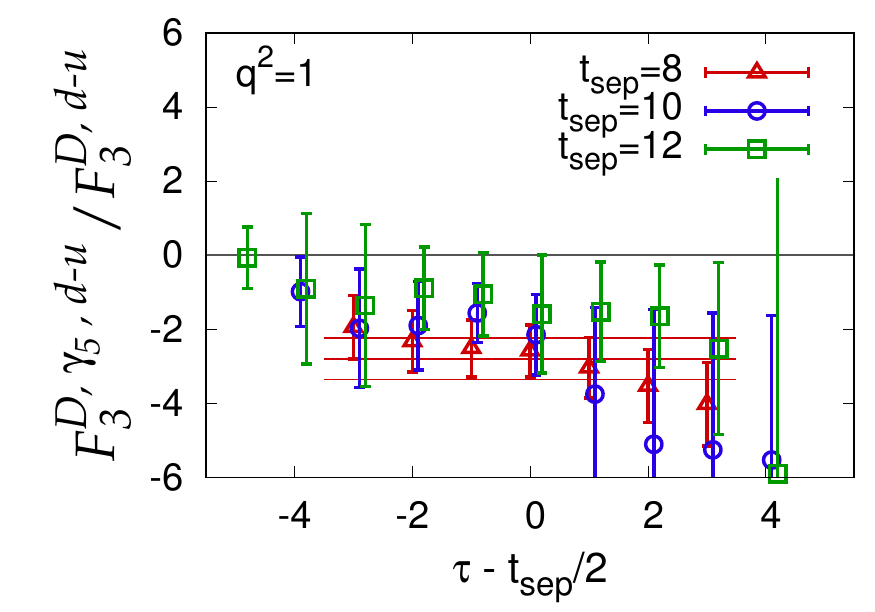}
  \includegraphics[width=0.24\textwidth]{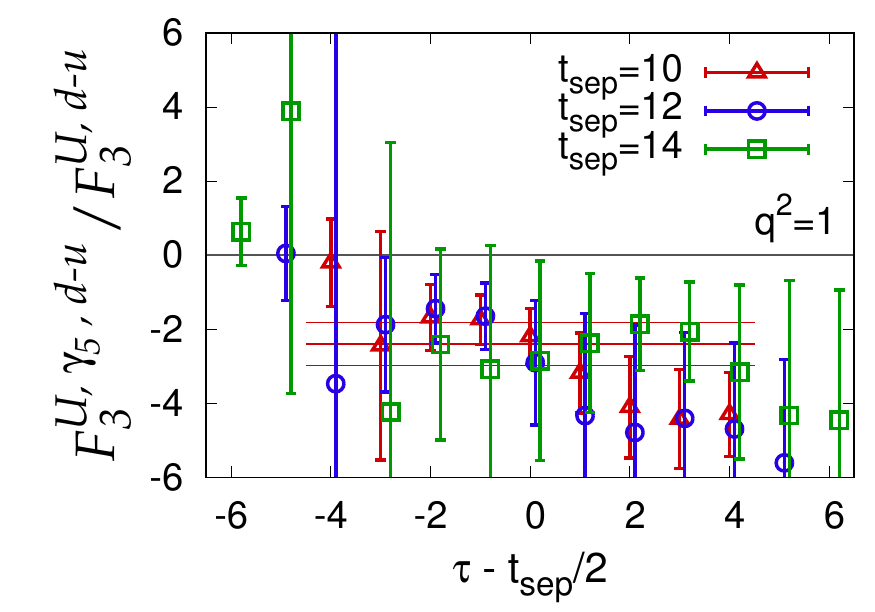}
  \includegraphics[width=0.24\textwidth]{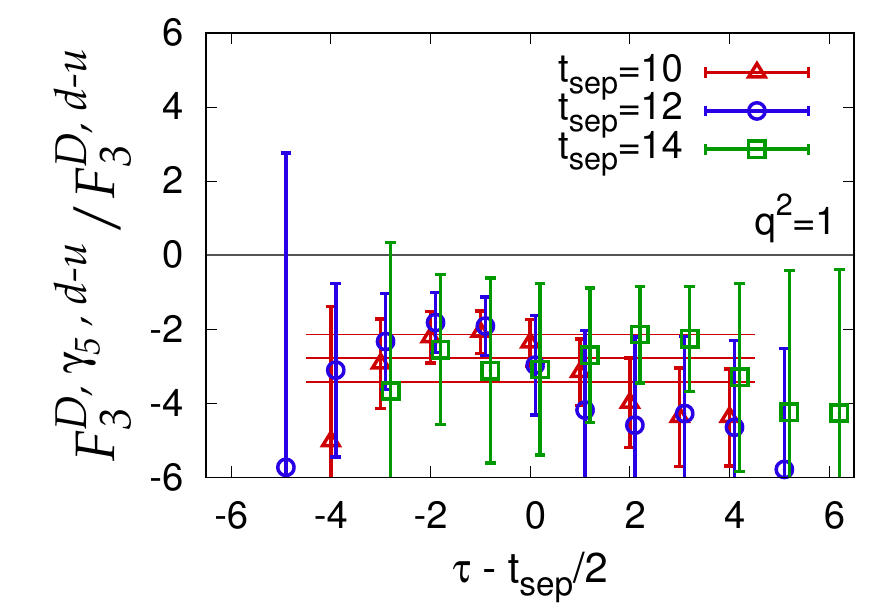}
}
{
  \includegraphics[width=0.24\textwidth]{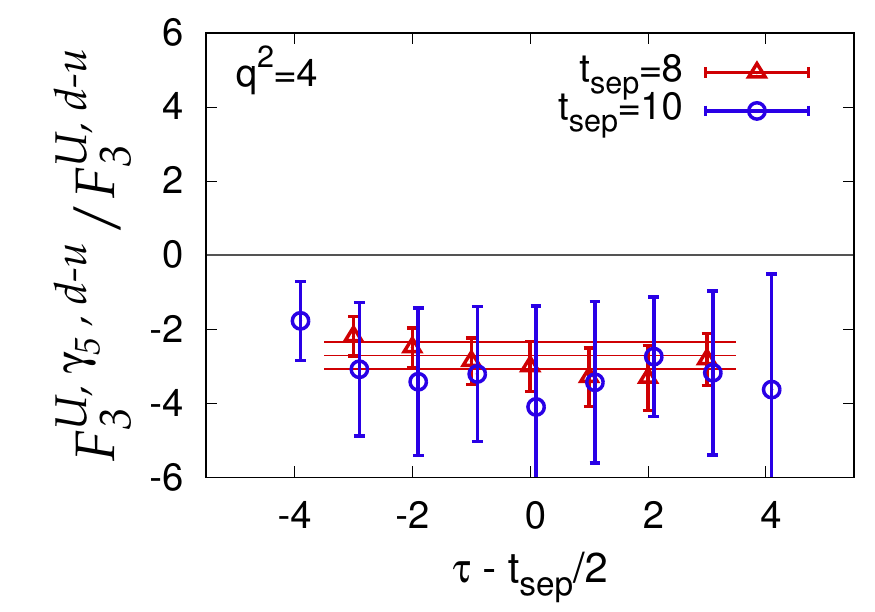}
  \includegraphics[width=0.24\textwidth]{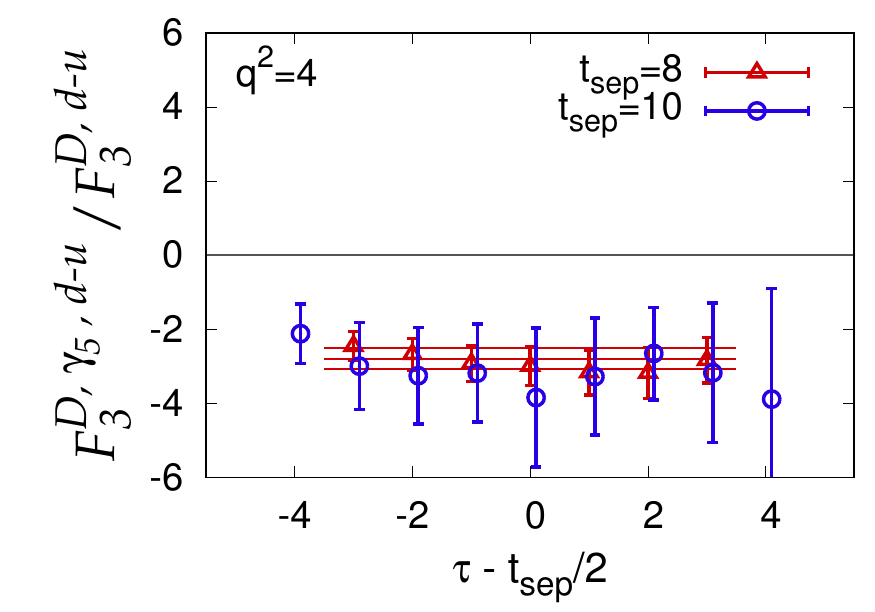}
  \includegraphics[width=0.24\textwidth]{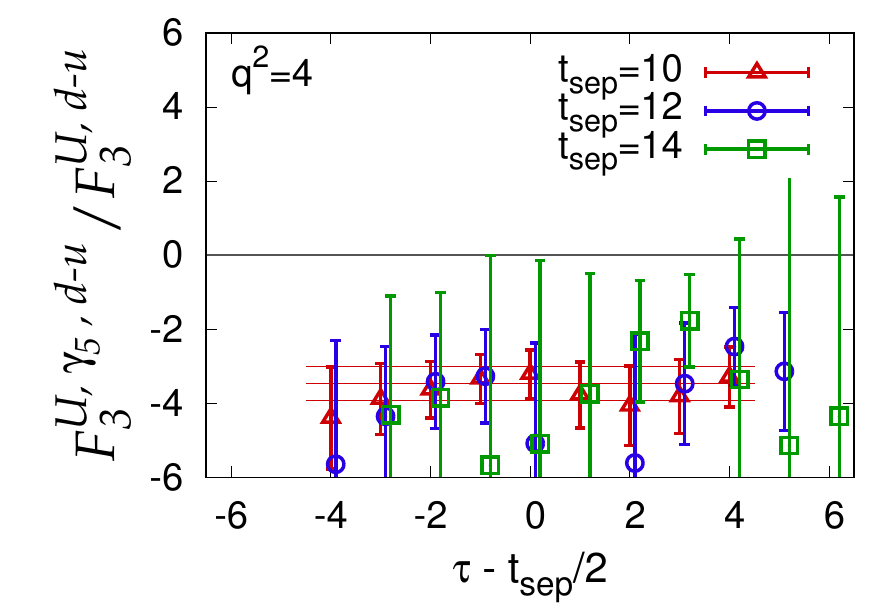}
  \includegraphics[width=0.24\textwidth]{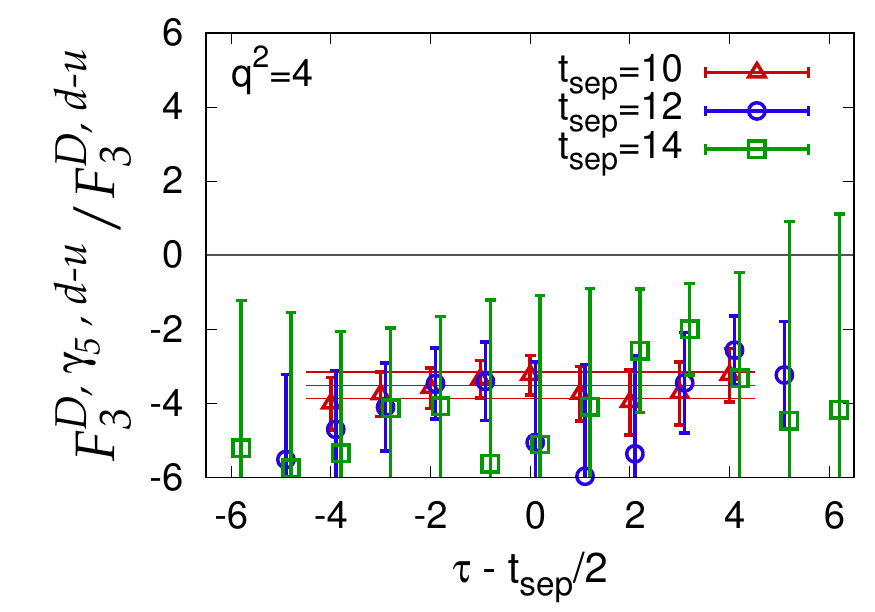}
}
  \caption{The ratio of the  connected contribution of the $\gamma_5$ term to the 
CEDM term in the $F_3$ form factor. The rest is the same as in Fig.~\protect\ref{fig:F3CEDM}.
  }
\label{fig:F3ratio}
\end{figure}

There are theoretical reasons to expect that the connected
contribution of the $\gamma_5$ operator is, with small corrections,
proportional to that of the CEDM operator~\cite{Bhattacharya:2016lat}.
In Fig.~\ref{fig:F3ratio}, we show the ratio of the two contributions
and find that this expectation is actually realized. 

To conclude, the data so far show that an acceptable signal-to-noise
ratio can be obtained in the connected contributions to $F_3$ for both
the CEDM and $\gamma_5$ operator insertion. We also find that The
ratio of the connected contribution of the $\gamma_5$ to CEDM term is
a constant.  This implies that the mixing of the $\gamma_5$ term with
the CEDM term can be cast as part of the multiplicative
renormalization of the CEDM operator. Controlling its $O(1/a^2)$ UV
divergent coefficient is a hurdle for lattice theories that respect
chiral symmetry (domain wall or overlap fermions) and those that don't
such as Wilson-clover fermions. Work to address the full mixing of the
CEDM operator and the calculation of the disconnected diagrams and the
reweighting factors is under progress.


\section*{Acknowledgments}
We thank the MILC Collaboration for providing the 2+1+1-flavor HISQ
ensembles. Simulations were carried out on computer facilities of (i)
Institutional Computing at Los Alamos National Lab; (ii) the USQCD
Collaboration, which are funded by the Office of Science of the
U.S. Department of Energy; (iii) the National Energy Research
Scientific Computing Center, a DOE Office of Science User Facility
supported by the Office of Science of the U.S. Department of Energy
under Contract No. DE-AC02-05CH11231.  The calculations used the
Chroma software suite~\cite{Edwards:2004sx}. Work supported by the
U.S. Department of Energy and the LANL LDRD program.

%
%
%

\end{document}